\documentclass[multphys,vecphys]{svmult}

\usepackage{makeidx}     
\usepackage{graphicx}    
\usepackage{multicol}    

\usepackage{cite}

\makeindex             

\begin{document}

\title*{Weak Universality of Spin Glasses\protect\newline
in Three Dimensions}

\titlerunning{Weak universality of 3D spin glasses}


\author{Tota Nakamura\inst{}
\and Shin-ichi Endoh\inst{}
\and Takeo Yamamoto\inst{}}

\authorrunning{T.~Nakamura et al.}

\institute{Department of Applied Physics, Tohoku University,
           Aoba-yama 05, 
Sendai, Miyagi 
980-8579, Japan
     }
%
%
\maketitle

\begin{abstract}
We point out a possibility of the weak universality of spin-glass
phase transitions in three dimensions.
The Ising, the XY, and the Heisenberg models with $\pm J$ bond distributions
undergo finite-temperature phase transitions with a ratio of
critical exponents $\gamma/\nu=2.4$.
Evaluated critical exponents agree with corresponding experimental results.
The analyses are based upon the nonequilibrium relaxation from the
paramagnetic state and the finite-time scaling.
\end{abstract}

%
%
%
%
%
%
%


\section{Introduction}
A spin glass (SG) phenomenon 
has been attracting great interests for thirty years
both theoretically and experimentally \cite{sgreview}.
It is also 
one of the toughest subject in a field of the computational physics.
Though many theoretical investigations have been done,
it is still an unsettled issue whether or not a simple random bond
model explains the SG phase transition.
Spins of many SG materials are well-approximated by the Heisenberg spins.
However, numerical studies have suggested that there is no
finite-temperature SG transition in the Heisenberg model \cite{mcmillan,olive}.
This is quite different from the Ising SG case:
many numerical investigations \cite{maricampbell} consistently explain
corresponding experimental results \cite{isingsgexp}.
Kawamura \cite{chirality} proposed the chirality mechanism
in order to settle the discrepancy in the Heisenberg spin glass.
The chiral-glass transition occurs without the spin-glass order.
A small but finite random anisotropy in the real materials
mixes the chirality and the spin.
This effect induces the observed SG transition.
The scenario sounds nice.
However, Matsubara \cite{matsubara1,matsubara2,matsubara3} 
recently recalculated the domain-wall
excess energy and the spin-glass susceptibility and suggested that
the finite SG transition occurs.
The methods are quite similar to the previous ones
\cite{mcmillan,olive}.
The final conclusion drastically changes by a subtle difference in the 
analyses of the obtained data.

Computer simulations on SG models suffer from serious slow dynamics.
The system sizes treated up to now are accordingly limited to very small ones, 
e.g.,  mostly a linear size $L=20$ or less in three dimensions.
The contradiction of the numerical results mentioned above
possibly originates in the smallness of the system.
This is a limit of the conventional equilibrium simulational study.
One realizes the equilibrium state in finite size systems.
Then, the thermodynamic properties are obtained
by the finite-size scaling analysis.

We take an opposite approach to the thermodynamic limit and
reexamine the SG transition.
First, we take the infinite size limit by dealing with a very large system
within a finite time range before the finite-size effect appears.
Then, the finite-time scaling analysis is performed to obtain the 
thermodynamic properties.
This method is known as the nonequilibrium relaxation (NER) method
\cite{ner,huse,blundell}.
In our previous paper \cite{totasg1} it is clarified that the SG transition
occurs at the same finite temperature as the chiral-glass transition occurs.
Of course, the chirality freezes if the spin freezes.
The estimated critical exponent $\gamma$ is consistent with the
corresponding experimental result \cite{heisenexp}.
The chirality mechanism is not necessary.
In this article we expand the same analysis to the Ising SG model and the XY SG
model in three dimensions.
It is found that the Ising, the XY, and the Heisenberg models 
undergo finite-temperature SG transitions with the weak universality.



%

\section{Method}
The model we work with is the nearest-neighbor $\pm J$ random bond model.
For each random bond configuration we prepare eight real replicas with
different paramagnetic spin configurations and with different random number
sequences.
Spins are updated by the single-spin-flip algorithm.
We calculate at each time step $t$ 
the spin-glass susceptibility $\chi_\mathrm{sg}$ through overlaps 
$q^{\alpha\beta}_{\mu, \nu}$ 
between two replicas ($\alpha$ and $\beta$) out of eight:
\[
\chi_\mathrm{sg}= 
\frac{1}{N}\sum_{i,j}\left[
\langle \mbox{\boldmath $S$}_i \cdot \mbox{\boldmath $S$}_j 
\rangle_\mathrm{therm} ^2
\right]_\mathrm{conf}
=
N
\left
[
\frac{2}{m(m-1)}\sum_{\alpha > \beta}^{m}\sum_{\mu,\nu} ^{x,y,z}
(q^{\alpha\beta}_{\mu,\nu})^2
\right 
] _\mathrm{conf}
\]
Indices $\mu$ and $\nu$ denote three components of spin degrees of freedom.
The thermal average is replaced by the summation over the real replicas,
whose number is denoted by $m$.
It determines the statistics of the thermal average.
Thus, a relaxation function $\chi_\mathrm{sg}(t)$ 
for one bond configuration is obtained.
A configurational average is taken over $\chi_\mathrm{sg}(t)$ with
different random bond configurations, 
different initial paramagnetic states, 
and different random number sequences.

The spin-glass susceptibility is expected to diverge at the 
transition temperature ($T_\mathrm{sg}$) as $\chi_\mathrm{sg}(t)\sim t^{\gamma/z\nu}$,
where $z$ is the dynamic exponent.
We obtain $T_\mathrm{sg}$, $\gamma$, and $z\nu$ by the
finite-time scaling analysis.
The NER of the Binder parameter $g(t)$ is also calculated.
Since the quantity is related to the fourth-order cumulant,
quite a lot of samples are necessary to obtain meaningful data.
Numbers of the bond configurations to obtain the results in this article
is summarized in Table \ref{tab:samplist}.
The Binder parameter is expected to behave at $T_\mathrm{sg}$
as $g(t)\sim t^{d/z}$, by which
$z$ is independently obtained.
Then, $\nu$ is estimated from a value of $z\nu$ obtained by the scaling.
The NER of the spin-glass susceptibility was originally applied to the
Ising SG model by Huse \cite{huse} and that of the
Binder parameter was applied by Blundell et al. \cite{blundell}.


\section{Results}

\begin{figure}[ht]
\centering
\includegraphics[width=5.7cm]{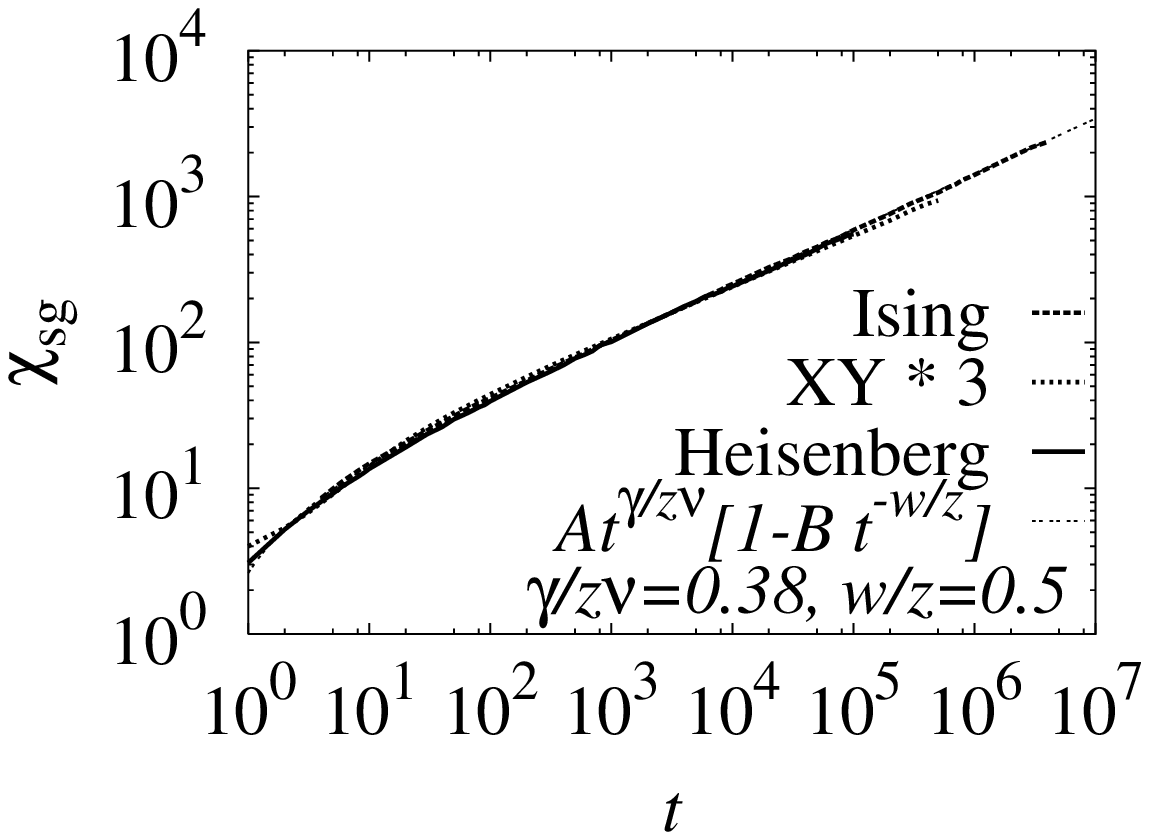}
\includegraphics[width=5.7cm]{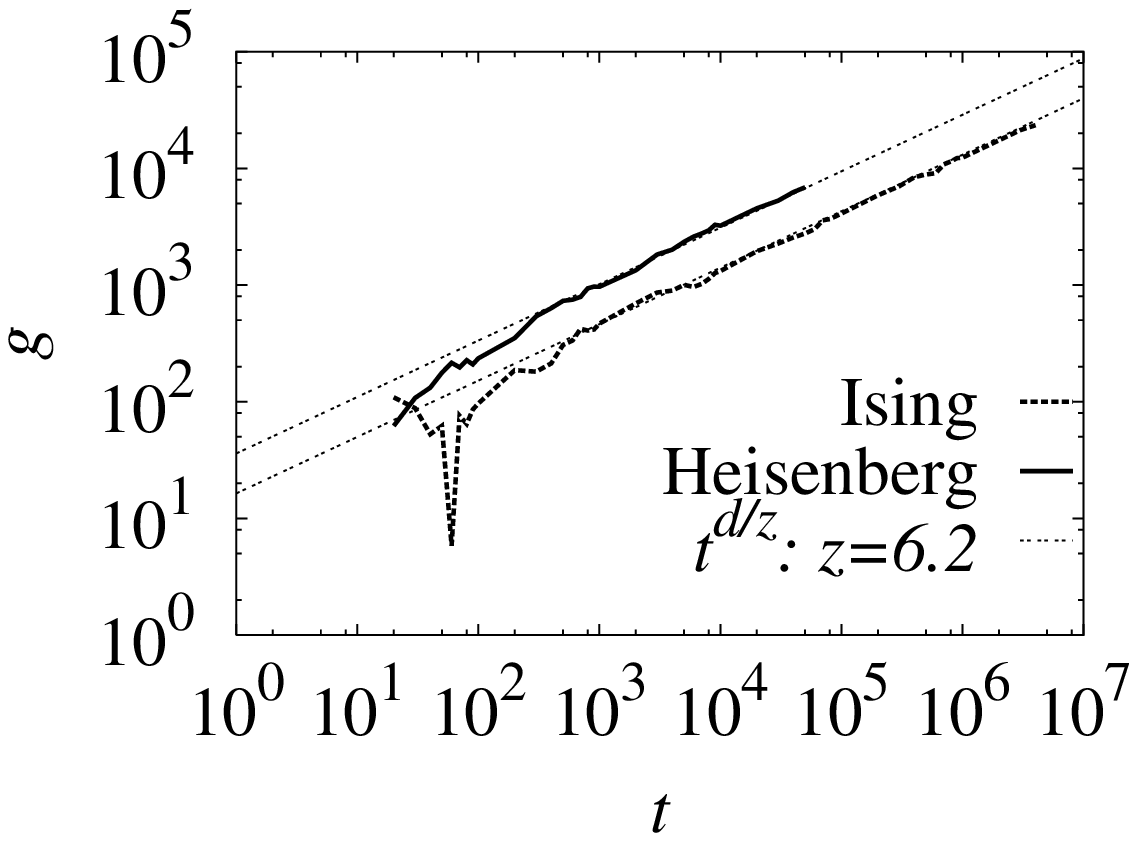}
\caption[]{({\bf Left}) NER of the $\chi_\mathrm{sg}$ at $T_\mathrm{sg}$.
Data of the XY model are multiplied by three.
The temperatures are
$T/J=1.17$ for the Ising model, 
$T/J=0.46$ for the XY model, and
$T/J=0.21$ for the Heisenberg model.
The linear size, time steps, and numbers of bond configurations 
are summarized in Table \ref{tab:samplist}.
({\bf Right}) NER of the Binder parameter 
for the Ising model and the Heisenberg model
}
\label{fig:weakuniv}
\end{figure}

Numerical results suggesting the weak universality is shown in 
Fig. \ref{fig:weakuniv}.
The NER data of the spin-glass susceptibility exhibit an algebraic
divergence with the exponent $\gamma/z\nu=0.38$.
The NER function of the Ising model and that of the Heisenberg model
are not distinguishable.
That of the XY model differs by a factor three.
An amplitude of the $\chi_\mathrm{sg}$ is dependent on the model.
If we take into account a correction-to-scaling term with its 
exponent $w/z=0.5$ \cite{maricampbell}, the relaxation functions fit to 
the expression $A t^{\gamma/z\nu}[1-B t^{-w/z}]$ from the first step.
A deviation of the XY data from the diverging line after $t=10^5$
is possibly due to the finite-size effect.
The NER of the Binder parameter shows an algebraic divergence with
$z=6.2$ for both models. 
Therefore, we obtain $\gamma/\nu=2.4$ which is common to the
Ising and the Heisenberg model.
This suggests the weak universality.
%


Results on the Heisenberg model are briefly reported in \cite{totasg1}.
We have obtained 
$T_\mathrm{sg}=0.20(2)$, $\gamma=1.9(5)$, and $z\nu=4.7(12)$.
Since $z=6.2$, $\nu$ is estimated as $\nu=0.8(2)$.
%
\begin{table}[ht]
\centering
\caption{Numbers of bond configurations to obtain $\chi_\mathrm{sg}$ and $g$
of Fig. \ref{fig:weakuniv}}
\begin{tabular}{clc cccccccc}
\hline
&&&time step&&&&&&\\
&model&size&$5\cdot 10^2$ &$1\cdot 10^3$&$1\cdot 10^4$&$5\cdot 10^4$&$1\cdot 10^5$&$5\cdot 10^5$&$1\cdot 10^6$&$4\cdot 10^6$\\
\hline
$\chi_\mathrm{sg}$& Ising&49&$\to$&$\to$&$\to$&$\to$&393&$\to$&$\to$&88\\
& XY&39&$\to$&$\to$&$\to$&$\to$&3149&120&&\\
& Heisenberg&89&$\to$&$\to$&58&$\to$&22&&&\\
\hline
$g$& Ising&39&$\to$&255480&85480&18576&12626&$\to$&1830&172\\
& Heisenberg&39&170414&43114&18316&7038&&&&\\
\hline
\end{tabular}
\label{tab:samplist}
\end{table}

\begin{figure}[ht]
\centering
\includegraphics[width=3.8cm]{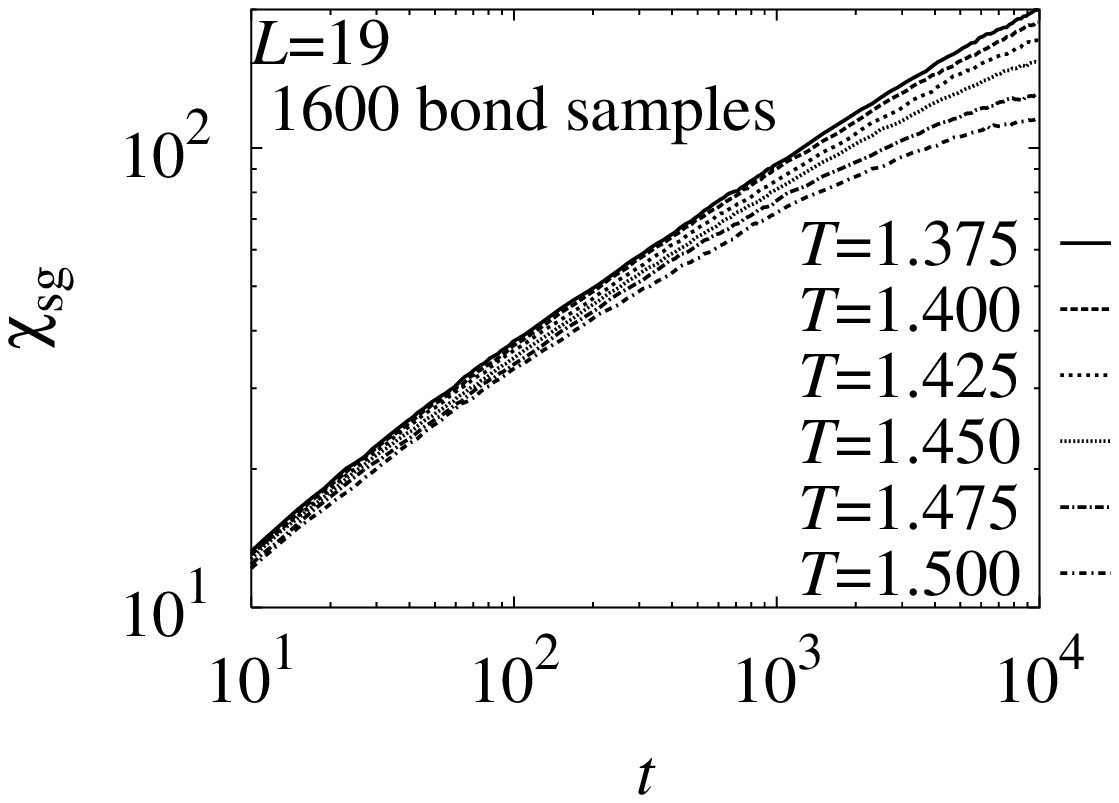}
\includegraphics[width=3.8cm]{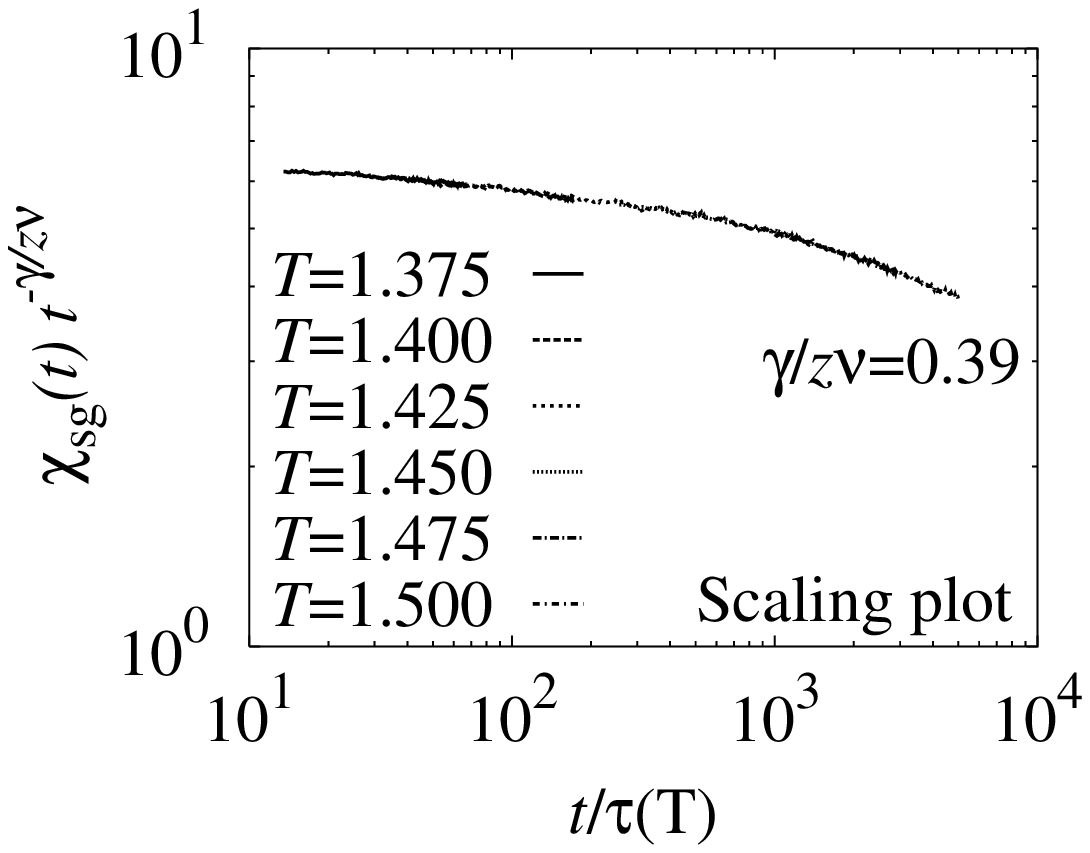}
\includegraphics[width=3.8cm]{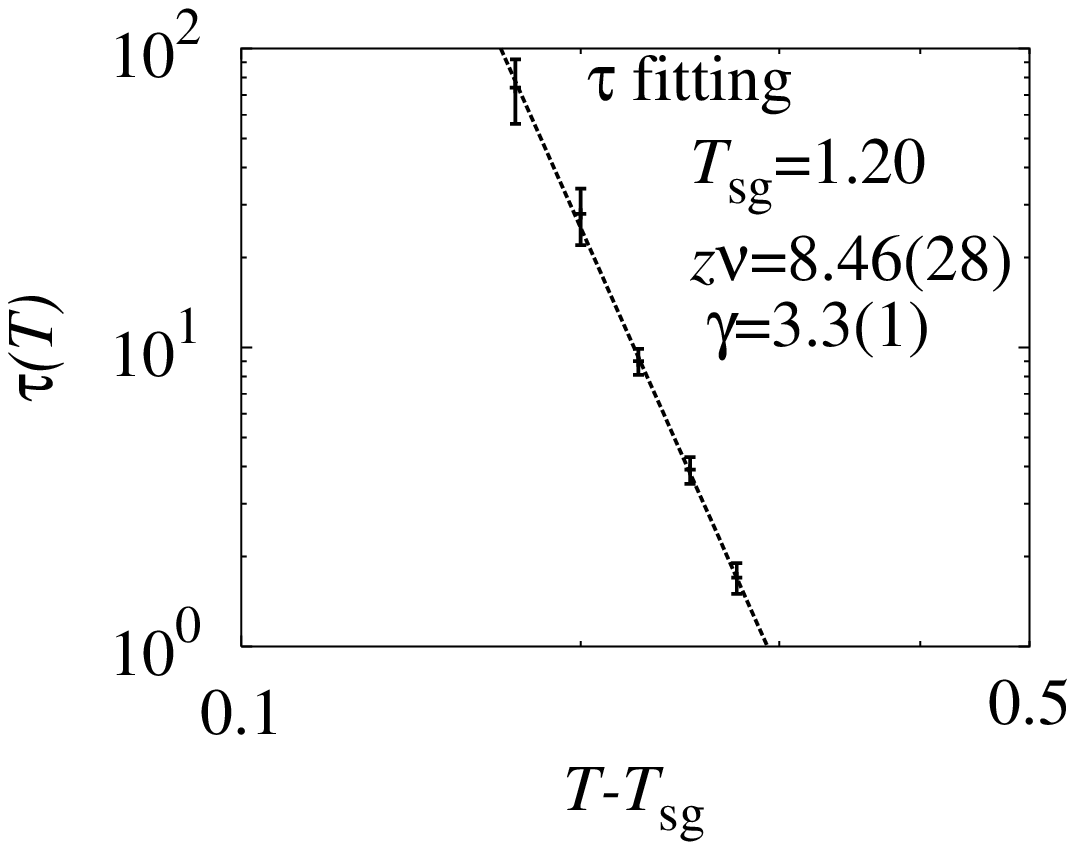}
\caption[]{({\bf Left}) NER of the $\chi_\mathrm{sg}$ of the Ising SG model
at high temperatures.
({\bf Middle})
A finite-time scaling plot.
We obtain $\tau(T)$ for each temperature and $\gamma/z\nu$ so that
the scaling is good.
It is also possible for $\gamma/z\nu=0.38\sim 0.40$.
({\bf Right})
Estimate $T_\mathrm{sg}$ supposing $\tau(T)\propto |T-T_\mathrm{sg}|^{-z\nu}$
}
\label{fig:isingsg}
\end{figure}

An example of a finite-time scaling analysis of the Ising model is shown in 
Fig. \ref{fig:isingsg}.
This is an only preliminary result with a small 
system size ($L=19$) and a short time range.
The error bars are consequently large.
Nevertheless, the obtained results are consistent with previous numerical
investigations \cite{maricampbell} and the corresponding experiment
\cite{isingsgexp} as summarized in Table \ref{tab:explist}.
The exponent estimated by the scaling 
($\gamma/z\nu=0.39$) is
consistent with that of the raw data at $T_\mathrm{sg}$ as
shown in Fig. \ref{fig:weakuniv}
($\gamma/z\nu=0.38$).

Time evolution of the distribution function of the overlap $P(q, t)$ is 
investigated at $T=T_\mathrm{sg}=1.17$ (figures not shown).
It exhibits a single Gaussian shape with a peak at $q=0$
before the size effect appears.
As the time increases, the width of the Gaussian distribution grows 
in accordance with the divergence of the spin-glass susceptibility.
It is possible to scale $P(q, t)/t^{\gamma/2z\nu}$ plotted versus
$qt^{\gamma/2z\nu}$ for various time steps from $t=10$ to $t=10^4$.
That is, the NER function knows the critical phenomenon from the 10th step.
The distribution changes its shape to having double peaks at $\pm q_\mathrm{eq}$
after the finite-size effect appears.

It is also possible to investigate the low-temperature phase.
The data suggest that the $\chi_\mathrm{sg}$ diverges algebraically 
with an exponent linearly dependent on the temperature.
Details will be reported elsewhere\cite{totaisingsg}.

\section{Concluding Remarks}

By applying the NER method it is clarified that the $\pm J$ spin glass models
in three dimensions undergo finite-temperature SG transitions.
There is possibly the weak universality
with a ratio of the critical exponents $\gamma/\nu=2.4$.
We compare our results with other numerical results and the experimental
results in Table \ref{tab:explist}.
They agree well within the numerical errors.
Critical exponents $\gamma$ and $\nu$ in the Heisenberg model are almost
one half of those in the Ising model.
Reexamination of the experiment on the critical exponents is strongly
encouraged.

The NER method is particularly effective in slow-dynamic systems.
Applications to various complex systems should be fruitful\cite{totaner}.
It potentially bring a breakthrough to a system stagnant due to 
the difficulty in computer simulations.

\begin{table}[ht]
\centering
\caption{Estimates of $T_\mathrm{sg}$, $\gamma$, 
$\nu$, $\beta$, $\gamma/\nu$, and $z$}
\begin{tabular}{l c c c c c c}
\hline
& $T_\mathrm{sg}$ & $\gamma$~ &  $\nu$ &$\beta$& $\gamma/\nu$ &$z$\\
\hline
Heisenberg SG\\
Present work 
& $0.20(2)$ & $1.9(5)$ & $0.8(2)$ & $-0.3\sim 0.7$ & $\sim 2.4$&$6.2$\\
Matsubara et al.\cite{matsubara3}
& $\sim 0.18$ & $\sim 2.06$ & $\sim 0.97$ & $\sim 0.43$ & $\sim 2.1$& \\

CdCr$_{2\times 0.85}$In$_{2\times 0.15}$S$_4$\cite{heisenexp}
& & $2.3(4)$ & $1.0\sim 1.5$ & $0.75(10)$ &$1.3 \sim 2.7$ & \\
\hline
Ising SG\\
Present work
& $1.17(4)$ & $3.6(6)$ & $1.5(3)$ & $\sim 0.5$ & $\sim 2.4$& $6.2$\\
Mari-Campbell \cite{maricampbell}
& $1.195(15)$ & $2.95(30)$ & $1.35(10)$ & $\sim 0.55$ & $\sim 2.2$ & \\
Fe$_{0.5}$Mn$_{0.5}$TiO$_3$\cite{isingsgexp}
&
& $4.0(3)$ & $\sim 1.7 $ & $\sim 0.54$ &$\sim 2.4$ & \\
\hline
\end{tabular}
\label{tab:explist}
\end{table}

\acknowledgement
The authors thank F.~Matsubara for fruitful discussions, and
N.~Ito and Y.~Kanada for providing them with
a random number generator.
Computations were partly done at the Supercomputer Center, 
ISSP, The University of Tokyo.
The author TN thanks a financial support from The
Sumitomo Foundation.


%

%

\end{document}